\documentclass[twocolumn,letterpaper,aps,prd,longbibliography,superscriptaddress,nofootinbib,floatfix]{revtex4-2}

\usepackage{graphicx}	
\usepackage{xspace}	

\begin{document}

\title{Transverse single-spin asymmetry of forward $\eta$ mesons \\
in $p^{\uparrow}$$+$$p$ collisions at $\sqrt{s}=200$ GeV}

\newcommand{\abilene}{Abilene Christian University, Abilene, Texas 79699, USA}
\newcommand{\augie}{Department of Physics, Augustana University, Sioux Falls, south Dakota 57197, USA}
\newcommand{\banaras}{Department of Physics, Banaras Hindu University, Varanasi 221005, India}
\newcommand{\barc}{Bhabha Atomic Research Centre, Bombay 400 085, India}
\newcommand{\baruch}{Baruch College, City University of New York, New York, New York, 10010 USA}
\newcommand{\bnlcoll}{Collider-Accelerator Department, Brookhaven National Laboratory, Upton, New York 11973-5000, USA}
\newcommand{\bnlphys}{Physics Department, Brookhaven National Laboratory, Upton, New York 11973-5000, USA}
\newcommand{\caucr}{University of California-Riverside, Riverside, California 92521, USA}
\newcommand{\charlesczech}{Charles University, Faculty of Mathematics and Physics, 180 00 Troja, Prague, Czech Republic}
\newcommand{\ciae}{Science and Technology on Nuclear Data Laboratory, China Institute of Atomic Energy, Beijing 102413, People's Republic of China}
\newcommand{\cns}{Center for Nuclear Study, Graduate School of Science, University of Tokyo, 7-3-1 Hongo, Bunkyo, Tokyo 113-0033, Japan}
\newcommand{\colorado}{University of Colorado, Boulder, Colorado 80309, USA}
\newcommand{\columbia}{Columbia University, New York, New York 10027 and Nevis Laboratories, Irvington, New York 10533, USA}
\newcommand{\czechtech}{Czech Technical University, Zikova 4, 166 36 Prague 6, Czech Republic}
\newcommand{\debrecen}{Debrecen University, H-4010 Debrecen, Egyetem t{\'e}r 1, Hungary}
\newcommand{\elte}{ELTE, E{\"o}tv{\"o}s Lor{\'a}nd University, H-1117 Budapest, P{\'a}zm{\'a}ny P.~s.~1/A, Hungary}
\newcommand{\eszterhazy}{Eszterh\'azy K\'aroly University, K\'aroly R\'obert Campus, H-3200 Gy\"ongy\"os, M\'atrai \'ut 36, Hungary}
\newcommand{\ewha}{Ewha Womans University, Seoul 120-750, Korea}
\newcommand{\fsu}{Florida State University, Tallahassee, Florida 32306, USA}
\newcommand{\gsu}{Georgia State University, Atlanta, Georgia 30303, USA}
\newcommand{\hanyang}{Hanyang University, Seoul 133-792, Korea}
\newcommand{\hiroshima}{Hiroshima University, Kagamiyama, Higashi-Hiroshima 739-8526, Japan}
\newcommand{\hunrenatomki}{HUN-REN ATOMKI, H-4026 Debrecen, Bem t{\'e}r 18/c, Hungary}
\newcommand{\ihepprot}{IHEP Protvino, State Research Center of Russian Federation, Institute for High Energy Physics, Protvino, 142281, Russia}
\newcommand{\illuiuc}{University of Illinois at Urbana-Champaign, Urbana, Illinois 61801, USA}
\newcommand{\inrras}{Institute for Nuclear Research of the Russian Academy of Sciences, prospekt 60-letiya Oktyabrya 7a, Moscow 117312, Russia}
\newcommand{\instpasczech}{Institute of Physics, Academy of Sciences of the Czech Republic, Na Slovance 2, 182 21 Prague 8, Czech Republic}
\newcommand{\isu}{Iowa State University, Ames, Iowa 50011, USA}
\newcommand{\jaea}{Advanced Science Research Center, Japan Atomic Energy Agency, 2-4 Shirakata Shirane, Tokai-mura, Naka-gun, Ibaraki-ken 319-1195, Japan}
\newcommand{\jeonbuk}{Jeonbuk National University, Jeonju, 54896, Korea}
\newcommand{\jyvaskyla}{Helsinki Institute of Physics and University of Jyv{\"a}skyl{\"a}, P.O.Box 35, FI-40014 Jyv{\"a}skyl{\"a}, Finland}
\newcommand{\kek}{KEK, High Energy Accelerator Research Organization, Tsukuba, Ibaraki 305-0801, Japan}
\newcommand{\korea}{Korea University, Seoul 02841, Korea}
\newcommand{\kurchatov}{National Research Center ``Kurchatov Institute", Moscow, 123098 Russia}
\newcommand{\kyoto}{Kyoto University, Kyoto 606-8502, Japan}
\newcommand{\labllr}{Laboratoire Leprince-Ringuet, Ecole Polytechnique, CNRS-IN2P3, Route de Saclay, F-91128, Palaiseau, France}
\newcommand{\lahorelums}{Physics Department, Lahore University of Management Sciences, Lahore 54792, Pakistan}
\newcommand{\lawllnl}{Lawrence Livermore National Laboratory, Livermore, California 94550, USA}
\newcommand{\losalamos}{Los Alamos National Laboratory, Los Alamos, New Mexico 87545, USA}
\newcommand{\lund}{Department of Physics, Lund University, Box 118, SE-221 00 Lund, Sweden}
\newcommand{\maryland}{University of Maryland, College Park, Maryland 20742, USA}
\newcommand{\mass}{Department of Physics, University of Massachusetts, Amherst, Massachusetts 01003-9337, USA}
\newcommand{\mate}{MATE, Institute of Technology, Laboratory of Femtoscopy, K\'aroly R\'obert Campus, H-3200 Gy\"ongy\"os, M\'atrai \'ut 36, Hungary}
\newcommand{\michigan}{Department of Physics, University of Michigan, Ann Arbor, Michigan 48109-1040, USA}
\newcommand{\miss}{Mississippi State University, Mississippi State, Mississippi 39762, USA}
\newcommand{\muhlenberg}{Muhlenberg College, Allentown, Pennsylvania 18104-5586, USA}
\newcommand{\myongji}{Myongji University, Yongin, Kyonggido 449-728, Korea}
\newcommand{\nagasaki}{Nagasaki Institute of Applied Science, Nagasaki-shi, Nagasaki 851-0193, Japan}
\newcommand{\nara}{Nara Women's University, Kita-uoya Nishi-machi Nara 630-8506, Japan}
\newcommand{\natmephi}{National Research Nuclear University, MEPhI, Moscow Engineering Physics Institute, Moscow, 115409, Russia}
\newcommand{\newmex}{University of New Mexico, Albuquerque, New Mexico 87131, USA}
\newcommand{\nmsu}{New Mexico State University, Las Cruces, New Mexico 88003, USA}
\newcommand{\northcg}{Physics and Astronomy Department, University of north Carolina at Greensboro, Greensboro, north Carolina 27412, USA}
\newcommand{\ohio}{Department of Physics and Astronomy, Ohio University, Athens, Ohio 45701, USA}
\newcommand{\ornl}{Oak Ridge National Laboratory, Oak Ridge, Tennessee 37831, USA}
\newcommand{\orsay}{IPN-Orsay, Univ.~Paris-Sud, CNRS/IN2P3, Universit\'e Paris-Saclay, BP1, F-91406, Orsay, France}
\newcommand{\pnpi}{PNPI, Petersburg Nuclear Physics Institute, Gatchina, Leningrad region, 188300, Russia}
\newcommand{\riken}{RIKEN Nishina Center for Accelerator-Based Science, Wako, Saitama 351-0198, Japan}
\newcommand{\rikjrbrc}{RIKEN BNL Research Center, Brookhaven National Laboratory, Upton, New York 11973-5000, USA}
\newcommand{\rikkyo}{Physics Department, Rikkyo University, 3-34-1 Nishi-Ikebukuro, Toshima, Tokyo 171-8501, Japan}
\newcommand{\saispbstu}{Saint Petersburg State Polytechnic University, St.~Petersburg, 195251 Russia}
\newcommand{\seoulnat}{Department of Physics and Astronomy, Seoul National University, Seoul 151-742, Korea}
\newcommand{\stonybrkc}{Chemistry Department, Stony Brook University, SUNY, Stony Brook, New York 11794-3400, USA}
\newcommand{\stonycrkp}{Department of Physics and Astronomy, Stony Brook University, SUNY, Stony Brook, New York 11794-3800, USA}
\newcommand{\sungskku}{Sungkyunkwan University, Suwon, 440-746, Korea}
\newcommand{\tenn}{University of Tennessee, Knoxville, Tennessee 37996, USA}
\newcommand{\titech}{Department of Physics, Tokyo Institute of Technology, Oh-okayama, Meguro, Tokyo 152-8551, Japan}
\newcommand{\tsukuba}{Tomonaga Center for the History of the Universe, University of Tsukuba, Tsukuba, Ibaraki 305, Japan}
\newcommand{\usmma}{United States Merchant Marine Academy, Kings Point, New York 11024, USA}
\newcommand{\vandy}{Vanderbilt University, Nashville, Tennessee 37235, USA}
\newcommand{\weizmann}{Weizmann Institute, Rehovot 76100, Israel}
\newcommand{\wigner}{Institute for Particle and Nuclear Physics, HUN-REN Wigner Research Centre for Physics, (HUN-REN Wigner RCP, RMI), H-1525 Budapest 114, POBox 49, Budapest, Hungary}
\newcommand{\yonsei}{Yonsei University, IPAP, Seoul 120-749, Korea}
\newcommand{\zagreb}{Department of Physics, Faculty of Science, University of Zagreb, Bijeni\v{c}ka c.~32 HR-10002 Zagreb, Croatia}
\newcommand{\zambia}{Department of Physics, School of Natural Sciences, University of Zambia, Great East Road Campus, Box 32379, Lusaka, Zambia}
\affiliation{\abilene}
\affiliation{\augie}
\affiliation{\banaras}
\affiliation{\barc}
\affiliation{\baruch}
\affiliation{\bnlcoll}
\affiliation{\bnlphys}
\affiliation{\caucr}
\affiliation{\charlesczech}
\affiliation{\ciae}
\affiliation{\cns}
\affiliation{\colorado}
\affiliation{\columbia}
\affiliation{\czechtech}
\affiliation{\debrecen}
\affiliation{\elte}
\affiliation{\ewha}
\affiliation{\fsu}
\affiliation{\gsu}
\affiliation{\hanyang}
\affiliation{\hiroshima}
\affiliation{\hunrenatomki}
\affiliation{\ihepprot}
\affiliation{\illuiuc}
\affiliation{\inrras}
\affiliation{\instpasczech}
\affiliation{\isu}
\affiliation{\jaea}
\affiliation{\jeonbuk}
\affiliation{\jyvaskyla}
\affiliation{\kek}
\affiliation{\korea}
\affiliation{\kurchatov}
\affiliation{\kyoto}
\affiliation{\labllr}
\affiliation{\lahorelums}
\affiliation{\lawllnl}
\affiliation{\losalamos}
\affiliation{\lund}
\affiliation{\maryland}
\affiliation{\mass}
\affiliation{\mate}
\affiliation{\michigan}
\affiliation{\miss}
\affiliation{\muhlenberg}
\affiliation{\myongji}
\affiliation{\nagasaki}
\affiliation{\nara}
\affiliation{\natmephi}
\affiliation{\newmex}
\affiliation{\nmsu}
\affiliation{\northcg}
\affiliation{\ohio}
\affiliation{\ornl}
\affiliation{\orsay}
\affiliation{\pnpi}
\affiliation{\riken}
\affiliation{\rikjrbrc}
\affiliation{\rikkyo}
\affiliation{\saispbstu}
\affiliation{\seoulnat}
\affiliation{\stonybrkc}
\affiliation{\stonycrkp}
\affiliation{\sungskku}
\affiliation{\tenn}
\affiliation{\titech}
\affiliation{\tsukuba}
\affiliation{\usmma}
\affiliation{\vandy}
\affiliation{\weizmann}
\affiliation{\wigner}
\affiliation{\yonsei}
\affiliation{\zagreb}
\affiliation{\zambia}
\affiliation{\wigner}
\author{N.J.~Abdulameer} \affiliation{\debrecen} \affiliation{\hunrenatomki}
\author{U.~Acharya} \affiliation{\gsu}
\author{C.~Aidala} \affiliation{\losalamos} \affiliation{\michigan} 
\author{N.N.~Ajitanand} \altaffiliation{Deceased} \affiliation{\stonybrkc} 
\author{Y.~Akiba} \email[PHENIX Spokesperson: ]{akiba@rcf.rhic.bnl.gov} \affiliation{\riken} \affiliation{\rikjrbrc}
\author{R.~Akimoto} \affiliation{\cns} 
\author{J.~Alexander} \affiliation{\stonybrkc} 
\author{D.~Anderson} \affiliation{\isu}
\author{S.~Antsupov} \affiliation{\saispbstu}
\author{K.~Aoki} \affiliation{\kek} \affiliation{\riken} 
% ...
\collaboration{PHENIX Collaboration}  \noaffiliation

\date{\today}


\begin{abstract}

Utilizing the 2012 transversely polarized proton data from the 
Relativistic Heavy Ion Collider at Brookhaven National Laboratory, the 
forward $\eta$-meson transverse single-spin asymmetry ($A_N$) was 
measured for $p^{\uparrow}+p$ collisions at $\sqrt{s}=200$ GeV as a 
function of Feynman-x ($x_F$) for $0.2<|x_F|<0.8$ and transverse 
momentum ($p_T$) for $1.0<p_T<5.0$ GeV/$c$. Large asymmetries at 
positive $x_F$ are observed ($\left<A_N\right>=0.086 \pm 0.019$), 
agreeing well with previous measurements of $\pi^{0}$ and $\eta$ $A_N$, 
but with reach to higher $x_F$ and $p_T$. The contribution of 
initial-state spin-momentum correlations to the asymmetry, as calculated 
in the collinear twist-3 framework, appears insufficient to describe the 
data.

\end{abstract}


\maketitle

\section{\label{sec:introduction}Introduction}

In quantum chromodynamics (QCD), the behavior of the theory depends 
strongly on the distance scale of the physics involved. At short 
distances where the strong coupling is small, the quarks and gluons, 
collectively known as partons, are readily described within the 
framework of perturbative QCD (pQCD). However, at long distances, the 
coupling becomes large and the theory becomes nonperturbative. 
Calculations of cross sections within QCD rely on factorizing the 
perturbative hard scattering at short distances from the nonperturbative 
physics at long distances, which is contained in the universal parton 
distribution functions (PDFs) and fragmentation functions 
(FFs)~\cite{Collins:1989gx}. The standard factorization procedure 
explicitly defines only the collinear degrees of freedom of the 
nonperturbative functions and then expands the cross section in inverse 
powers of the momentum transfer $Q$, alongside the usual expansion of 
the hard scattering in powers of the strong coupling, $\alpha_s$.

Historically, unpolarized observables, such as inclusive cross sections, 
have been well described by this collinear QCD factorization scheme at 
the leading $n=2$ term of the $1/Q^{n-2}$ factorization expansion, where 
$n$ is known as the twist. However, large azimuthal asymmetries in 
particle production from transversely polarized proton collisions 
($p^{\uparrow}+p \rightarrow h + 
X$)~\cite{ADAMS1991201,ADAMS19983,PhysRevLett.101.222001,PhysRevD.86.051101,PhysRevD.90.012006}, 
particularly at forward rapidities, disagree with the leading-twist 
collinear pQCD predictions of asymmetries on the order of 
$10^{-4}$~\cite{PhysRevLett.41.1689}. These transverse single-spin 
asymmetries (TSSAs) persist at high 
$p_T$~\cite{PhysRevLett.101.222001,PhysRevD.90.012006,PhysRevD.103.072005} 
where pQCD is valid, suggesting additional considerations are needed in 
the nonperturbative initial- and final-state functions.

Multiple factorization frameworks have since been proposed with 
mechanisms to generate TSSAs. Transverse-momentum-dependent (TMD) 
factorization generalizes the definition of the PDFs and FFs to include 
explicit dependence on a nonperturbative transverse momentum. The 
resulting TMD distributions describe QCD spin-momentum correlations 
between the partons and hadrons. Two polarized TMD distributions have 
been identified as sources for large TSSAs. In the initial state, the 
Sivers TMD PDF~\cite{PhysRevD.41.83,PhysRevD.43.261} correlates the 
transverse polarization of the hadron with the transverse momentum of a 
constituent parton and explains the left-right asymmetry in particle 
production as arising from a transverse momentum imbalance of the 
partons. In the final state, the Collins TMD FF~\cite{COLLINS1993161} 
correlates the transverse polarization of the struck parton and the 
angular distribution of the fragmenting hadrons. The Collins effect 
attributes the left-right asymmetries to the fragmentation of 
transversely polarized quarks with a preferred transverse direction. It 
arises from the coupling of the Collins TMD FF with the quark 
transversity distribution~\cite{PhysRevLett.67.552,BARONE20021}, which 
encodes the correlation between the transverse polarization of the 
initial-state proton and that of its constituent quarks.

Experimental probes of TMDs require observables with access to both a 
hard scale $Q$ and a soft scale that defines the nonperturbative 
transverse momentum component, $k_T \ll Q$. In semi-inclusive 
deep-inelastic scattering (SIDIS), measurements of both pion and kaon 
single-spin 
asymmetries~\cite{Airapetian_2005,Alexakhin_2005,Alekseev_2009,Airapetian_2009} 
led to the extraction of the first nonzero Sivers functions for $u$ and 
$d$ 
quarks~\cite{PhysRevD.71.074006,PhysRevD.72.054028,Anselmino_2008,PhysRevD.73.014021}. 
Subsequently, nonzero Collins FFs for pions and transversity 
distributions for $u$ and $d$ quarks were 
extracted~\cite{PhysRevD.75.054032,PhysRevD.87.094019,PhysRevD.93.014009} 
from a combination of 
SIDIS~\cite{AGEEV200731,AIRAPETIAN201011,ADOLPH2012376} and $e^+e^-$ 
data~\cite{PhysRevD.78.032011}. Recent measurements have confirmed 
significant Collins asymmetries at much higher 
precision~\cite{PhysRevD.90.052003,PhysRevLett.116.042001,PhysRevD.100.092008}.

For processes with one accessible hard scale, such as inclusive hadron 
production at high $p_T$ in hadronic collisions, TSSAs can be described 
in collinear factorization with higher-twist multiparton 
correlators~\cite{EFREMOV1985383,PhysRevD.59.014004,Kanazawa_2000}. 
These twist-3 correlators represent the quantum interference between 
scattering with a single active parton and a composite parton state that 
contains an additional gluon field. They have been shown to be related 
to the $k_T$ moments of the TMD PDFs and 
FFs~\cite{PhysRevLett.97.082002}. Asymmetries of inclusive hadrons from 
hadronic collisions can receive contributions from correlators in both 
the initial and final state. Previous phenomenological studies attempted 
to describe $\pi^0$ and $\eta$ TSSAs in hadronic collisions with the 
initial-state Sivers-like twist-3 
correlator~\cite{PhysRevD.82.034009,Kanazawa:2011bg}. More recent 
studies indicate that the final-state Collins-like twist-3 term may be 
of considerable importance~\cite{Metz_2013,PhysRevD.89.111501} compared 
to other higher-twist terms.

Additional sources for TSSAs have been suggested, including diffractive 
processes~\cite{STAR:2020nnl} and nonperturbative color-exchange effects 
that may signal factorization breaking~\cite{Rogers:2010dm, 
PhysRevD.102.114012}. However, the extent to which such mechanisms 
contribute to inclusive forward hadron asymmetries in hadronic 
collisions is not yet well established. Thus, it remains an open 
question which combination of initial-state and final-state twist-3 
correlators, or alternative contributions, dominates the generation of 
the observed asymmetries.

In this article, the PHENIX collaboration reports the measurement of the 
forward $\eta$-meson TSSAs from $\sqrt{s}=200$ GeV $p^{\uparrow}+p$ 
collisions recorded in 2012. The TSSAs expand the kinematic reach on a 
previous $\sqrt{s}=200$ GeV PHENIX measurement from the 2008 
dataset~\cite{PhysRevD.90.072008} to higher $x_F$ and $p_T$. This 
previous analysis at $\sqrt{s}=200$ GeV also included a cross section 
measurement which exhibited good agreement with pQCD predictions, 
confirming the validity of factorization in this regime. We combine the 
measurements from both years of data taking and compare the resulting 
TSSAs to collinear twist-3 predictions of the asymmetry generated by 
only the initial-state Sivers-like contribution. 
It is also compared to the $\pi^0$ TSSAs in a similar rapidity range, 
which can illuminate any potential dependence of the asymmetry on 
isospin, strange quarks, and mass.

\section{\label{sec:experiment}The PHENIX experiment at RHIC}

\subsection{\label{sec:rhic}RHIC polarized proton beams}

The Relativistic Heavy Ion Collider (RHIC) at Brookhaven National 
Laboratory collides beams of polarized protons up to center of mass 
energies of 510 GeV. The two collider rings, called yellow and blue, 
store transversely polarized protons in 111 bunches with 
$\approx$~$10^{11}$ protons per bunch. Depolarizing resonances in the 
rings are mitigated through the implementation of two Siberian snakes, 
which rotate the proton polarization 180$^{\circ}$ without incurring 
orbit distortions~\cite{ALEKSEEV2003392}.

The beam polarization $P$ for (vertical) transversely polarized protons 
is defined as the difference in the number of protons with spin aligned 
versus anti-aligned with the vertical axis divided by the sum. Two 
separate polarimeters measure $\left<P\right>$ throughout the run for 
each beam using the asymmetry of hadronic elastic scattering in the 
Coulomb-nuclear-interference region~\cite{ALEKSEEV2003392}. A 
proton-carbon polarimeter makes quick measurements of the relative 
(uncalibrated) polarization several times within a 
store~\cite{Nakagawa:2008zzb}. They are normalized by the absolute 
polarization which is determined using a hydrogen gas jet 
polarimeter~\cite{PhysRevD.79.094014}. The absolute polarization 
measurements can take up to several hours due to statistical limitations 
arising from the low density of the hydrogen gas. Typically, the beam 
polarizations at RHIC range from $\approx$~50\% to 60\% with an 
uncertainty of 4\%--7\% for a few-hour store.

In each store during transverse-polarization running, the polarization 
direction of the proton bunches is alternated according to a 
predetermined pattern. Systematic effects stemming from bunch-by-bunch 
luminosity differences or detector efficiencies can be mitigated by 
having protons in both beams polarized in both directions. Because a 
TSSA measurement relies on the collision of a transversely polarized 
proton with an unpolarized proton, the sensitivity to polarization is 
removed in one beam by integrating over both spin-up and spin-down 
bunches.

\subsection{\label{sec:phenix}PHENIX forward detectors}

The PHENIX experiment at RHIC~\cite{ADCOX2003469} utilized a versatile 
detector apparatus, purpose-built for the high-resolution measurement of 
leptons, photons, and charged hadrons at the central rapidity of 
$|\eta|<0.35$, and with additional capabilities for the measurement of 
charged hadrons and muons at $1.2<|\eta|<2.2$. In the forward region, 
PHENIX consisted of an electromagnetic calorimeter known as the Muon 
Piston Calorimeter (MPC) and two global detectors: the beam-beam counter 
(BBC) and the zero-degree calorimeter (ZDC).

Located 220~cm along the beamline on either side of the nominal PHENIX 
interaction point and covering $3.1<|\eta|<3.9$, the MPC was a 
scintillating electromagnetic calorimeter comprising 220~(196) PbWO$_4$ 
crystal towers in its north (south) arms. Previous measurements of cross 
sections and asymmetries have exhibited the capability of the MPC to 
reconstruct light neutral mesons in the forward region through their 
decays to 
photons~\cite{PhysRevD.90.012006,PhysRevD.90.072008,PHENIX:2025sob}. To 
sample high-momentum $\eta$-meson candidates, an MPC trigger fired when 
a localized 4-by-4 group of towers had a cumulative energy deposition 
greater than a set threshold. In $\sqrt{s}=200$ GeV collisions, this 
threshold corresponded to cluster energies of roughly 20 GeV.

The BBC~\cite{ALLEN2003549}, located at $3.0<|\eta|<3.9$, had two 
arms of quartz \v{C}erenkov radiator arrays 144 cm to the north and south 
side of the nominal PHENIX collision point. It was used primarily to 
define the condition of a minimum-bias (MB) event and to provide a vertex 
position along the beam axis by utilizing its precise timing resolution 
to measure the difference in the particle time of flight between the 
north and south arms. The ZDC~\cite{ALLEN2003549} was a hadronic 
calorimeter in the far-forward region ($\eta \gtrsim 6$) used for 
neutron detection from diffractive or double-diffractive events. By 
sandwiching scintillating hodoscopes known as the shower-maximum 
detectors (SMD) within the ZDCs, PHENIX was also sensitive to the 
position of these neutrons with $\approx$~1~cm resolution. The ZDC/SMD 
system functioned as a local polarimeter for PHENIX through the 
measurement of known neutron single-spin asymmetries~\cite{Fukao_2007}. 
In 2012, results from the PHENIX local polarimeter confirmed the nominal 
vertical polarization direction of both beams.

\section{\label{sec:methods}Analysis Methods}

In transversely polarized $p^{\uparrow}$$+$$p$ collisions, the cross 
section of inclusive particle production is modified to first order by 
an azimuthal-cosine modulation known as the raw asymmetry,
\begin{equation} \label{eq:4}
\epsilon(\phi) = P A_N \cos{\phi} ,
\end{equation}
where $P$ is the beam polarization and $A_N$ is the TSSA. $A_N$ is 
measured by fitting a cosine function to the raw asymmetry and dividing 
the amplitude of the fit by the beam polarization. It quantifies the 
difference between azimuthally dependent cross sections when the proton 
is polarized up $(d\sigma^{\uparrow})$ versus down 
$(d\sigma^{\downarrow})$
\begin{equation} \label{eq:5}
A_N \cos{\phi} = \frac{1}{P}\frac{d\sigma^{\uparrow}(\phi) - 
d\sigma^{\downarrow}(\phi)}{ d\sigma^{\uparrow}(\phi) + 
d\sigma^{\downarrow}(\phi)},
\end{equation}
or, equivalently, when the final-state particle is produced to the left 
or right with respect to the polarized proton going direction. The angle 
$\phi$ is measured in the vertical plane, with the up direction being 
$\phi=\pi/2$. By convention, to the left~(right) of the polarized proton 
going direction is defined as $\phi=0\left(\phi=\pi\right)$. Rotating in 
this plane clearly gives $d\sigma^{\uparrow}(\phi) = 
d\sigma^{\downarrow}(\phi+\pi)$ and $d\sigma^{\downarrow}(\phi) = 
d\sigma^{\uparrow}(\phi+\pi)$.

During the 2012 data taking period, the south MPC underwent an upgrade, 
leaving only the north MPC to collect data for this measurement. It 
collected 1.23 $\times$ 10$^{-2}$ pb$^{-1}$ of MB data and 
12.95 pb$^{-1}$ of MPC triggered data. By treating the north-going 
(blue) beam as polarized and south-going (yellow) beam as unpolarized, 
$A_N$ can be measured as a function of positive $x_F=2p_z/\sqrt{s}$. 
Likewise, a polarized yellow and unpolarized blue beam provides access 
to negative $x_F$ values. As such, $A_N$ is measured at both positive 
and negative $x_F$ with bin widths of 0.1 from 0.2 to 0.4 and -0.2 to 
-0.4 in the MB data and 0.3 to 0.8 and -0.3 to -0.8 in the 
MPC-triggered data. Additionally, $A_N$ is measured as a function of 
$p_T$ in MB from 1.0 to 2.5 GeV/$c$ and with MPC triggers from 
2.0 to 5.0 GeV/$c$ in $p_T$ bins of width 0.5 GeV/$c$. The high-$x_F$ 
behavior of the asymmetry is explored further by separating $A_N$ versus 
$p_T$ into $x_F$ regions of 0.2 to 0.6 and 0.6 to 0.8.

\subsection{\label{sec:pol}Polarization}

The polarizations of both RHIC beams were tracked throughout the 2012 
data taking period using the proton-carbon and hydrogen gas jet 
polarimeters introduced in section~\ref{sec:rhic}. The 
luminosity-weighted magnitude of the proton polarization was determined 
to be $\left<P_{\rm blue}\right>=0.652 \pm 0.009$ (stat) $\pm$ 0.022 (syst) 
and $\left<P_{\rm yellow}\right>=0.587 \pm 0.009$ (stat) $\pm$ 0.020 (syst). 
The ratio in luminosity between polarized-up and polarized-down bunches, 
known as the relative luminosity $\mathcal{R}$, is an important quantity 
when determining the raw asymmetry. It is calculated using 
bunch-dependent trigger counts from the BBC. For both the blue and 
yellow beams in 2012, $\mathcal{R} \approx$~1 within a tenth of a percent.

\subsection{\label{sec:rawasyms}Raw asymmetries}

\begin{figure}[hbt]
\includegraphics[width=1.0\linewidth]{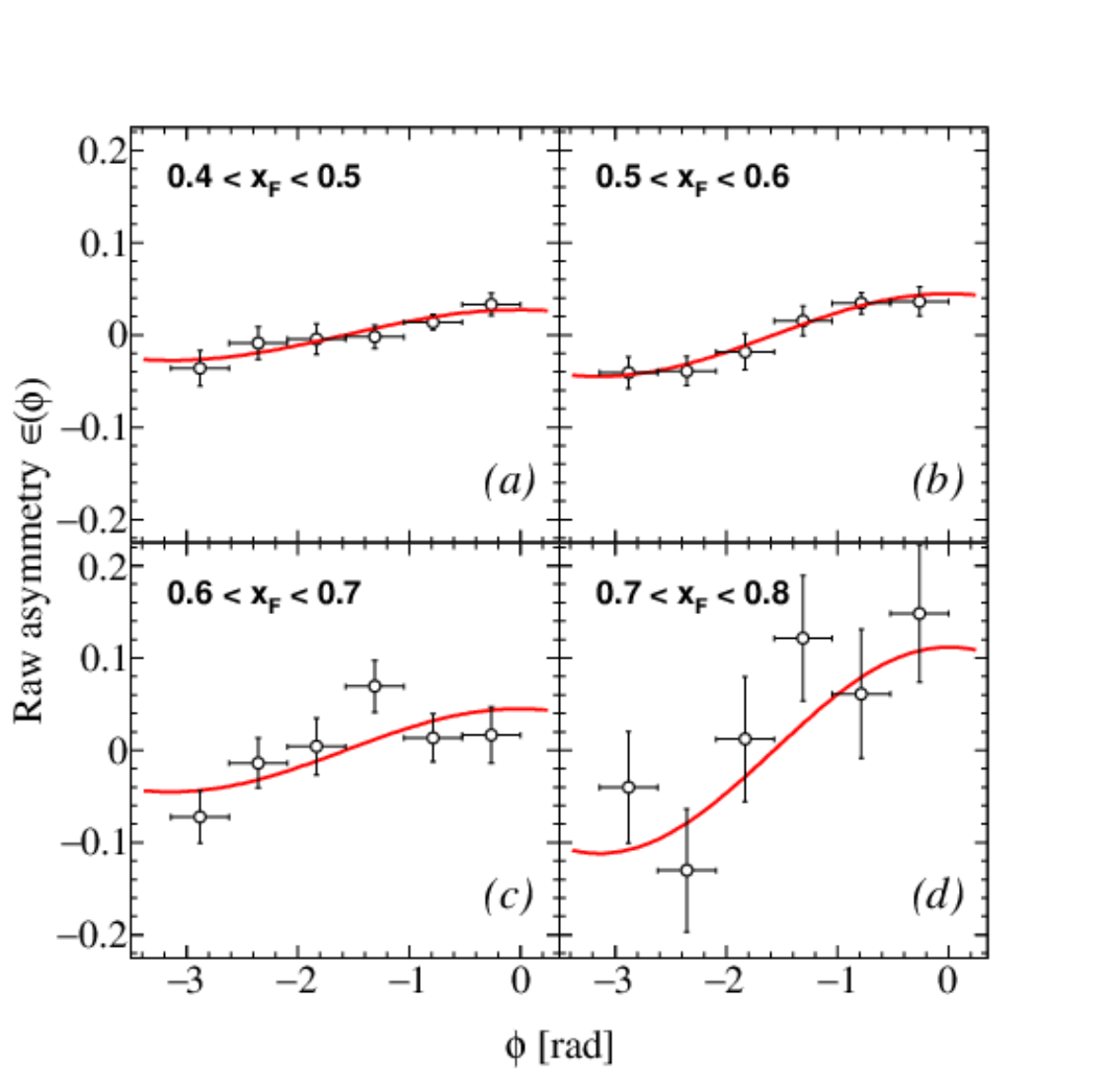}
\caption{\label{fig:rawasyms}A selection of MPC-triggered raw 
asymmetries and their cosine fits in the $p_T$ range between 1.0 and 5.0 
GeV/$c$ for (a) $0.4 < x_F < 0.5$ (b) $0.5 < x_F < 0.6$ (c) $0.6 < x_F < 
0.7$ (d) $0.7 < x_F < 0.8$.}
\end{figure}

\begin{figure*}[ht!]
    \includegraphics[width=0.48\linewidth]{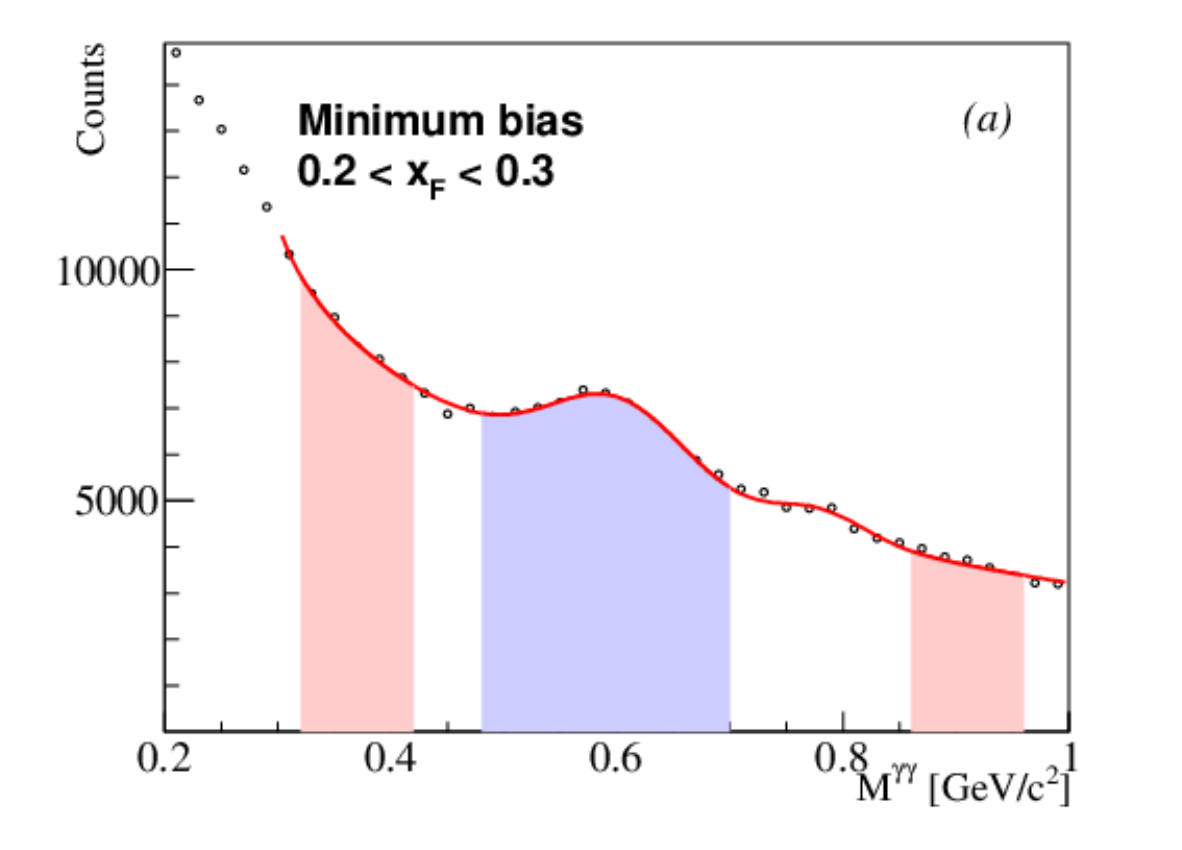}
    \includegraphics[width=0.48\linewidth]{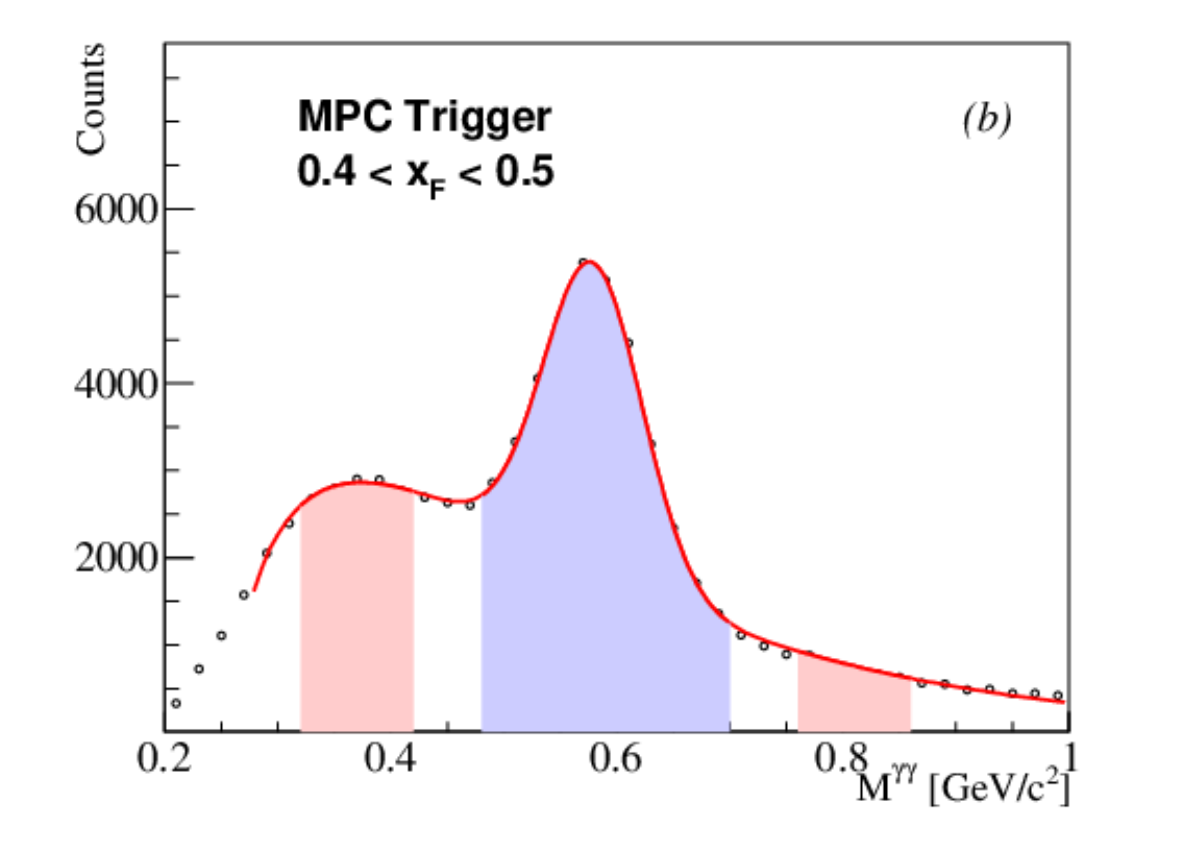}
  \caption{Examples of two-photon invariant mass distributions in the 
MPC at $\sqrt{s} = 200$ GeV from the 2012 data taking period in (a) 
MB and (b) MPC-triggered data. The blue filled regions are 
used for the $\eta$ meson-peak asymmetries while the red filled regions 
are designated as the sidebands for calculation of the background 
asymmetries.}
  \label{fig:invmass_sidebands}
\end{figure*}

Two independent methods are used to extract the raw asymmetry, 
$\epsilon(\phi)$. The first utilizes a geometric mean of yields on the 
left and right sides of the MPC from collisions of up and down polarized 
protons (Fig.~\ref{fig:rawasyms}),
\begin{equation} \label{eq:6}
\epsilon_{}(\phi) = \frac{\sqrt{N^{\uparrow}(\phi) N^{\downarrow}(\phi + \pi)} - \sqrt{N^{\downarrow}(\phi) N^{\uparrow}(\phi + \pi)}}{\sqrt{N^{\uparrow}(\phi) N^{\downarrow}(\phi + \pi)} + \sqrt{N^{\downarrow}(\phi) N^{\uparrow}(\phi + \pi)}} ,
\end{equation}

\noindent where the yields $N^{\uparrow,\downarrow}$ are determined by 
integrating the invariant mass distribution of diphoton candidates from 
the decay $\eta \rightarrow \gamma \gamma$ in the signal region of 0.48 
$< M_{\gamma\gamma} < $ 0.70 GeV/$c$$^2$ 
(Fig.~\ref{fig:invmass_sidebands}). This formula, to first order, 
cancels detector efficiency and relative luminosity effects. As a cross 
check, the raw asymmetry can also be measured with a formula that relies 
on counts from the same azimuthal region,
\begin{equation} \label{eq:7}
\epsilon(\phi) = \frac{N^{\uparrow}(\phi) - \mathcal{R}N^{\downarrow}(\phi)}{N^{\uparrow}(\phi) + \mathcal{R}N^{\downarrow}(\phi)} .
\end{equation}
This method must account for the relative luminosity between up and down 
bunches. The final $A_N$ is determined with Eq.~\ref{eq:6} while the 
difference between the two methods is taken as an uncorrelated 
systematic uncertainty.

\subsection{\label{sec:bkg}Background correction}

The asymmetry in the signal region must undergo a purity correction to 
account for the background underneath the $\eta$ meson peak. The 
presence of a combinatorial background in the signal region dilutes the 
asymmetry. In contrast, correlated background clusters, composed 
primarily of $\pi^0$ decay photons that are reconstructed in the same 
cluster due to their small opening angle (merged $\pi^0$ clusters), 
could generate some false asymmetry in the $\eta$ meson mass window that 
enhances the signal $A_N$. Both effects can be corrected for the TSSA 
$(A_N)$ as
\begin{equation} \label{eq:8}
A_N = \frac{A_N^{\rm peak} - rA_N^{\rm bkg}}{1-r} ,
\end{equation}
where $r = N_{\rm bkg}/(N_{\eta}+N_{\rm bkg})$ is the fraction of 
background in the signal region, $A_N^{\rm peak}$ is the transverse 
single-spin asymmetry in the signal region, and $A_N^{\rm bkg}$ is the 
transverse single-spin asymmetry in the sideband mass regions.

The background fraction is determined by fitting the invariant mass 
spectra with a gamma distribution background and Gaussian signal, as 
seen in Fig.~\ref{fig:invmass_sidebands}. In the MB data, an additional 
Gaussian centered at 0.78 GeV/$c$$^{2}$ was included in the background 
fit function to account for the $\omega(782) \rightarrow \pi^0 + \gamma$ 
decay. This $\omega(782)$ signal does not appear in the MPC 4-by-4 
triggered sample because the merged $\pi^0$ and $\gamma$ typically do 
not have enough energy to exceed the trigger threshold by themselves and 
the opening angle at $\sqrt{s}=200$ GeV is such that they rarely are 
found in the same 4-by-4 tile. In the MB (MPC-triggered) data, the 
background fractions are $\approx$~75\%~(55\%).
A systematic uncertainty is taken as the difference between background 
fractions from the nominal extraction using a functional fit and a 
separate estimation using Gaussian-process regression~\cite{Rasmussen2006}.

\begin{figure}[hb!]
\includegraphics[width=1.0\linewidth]{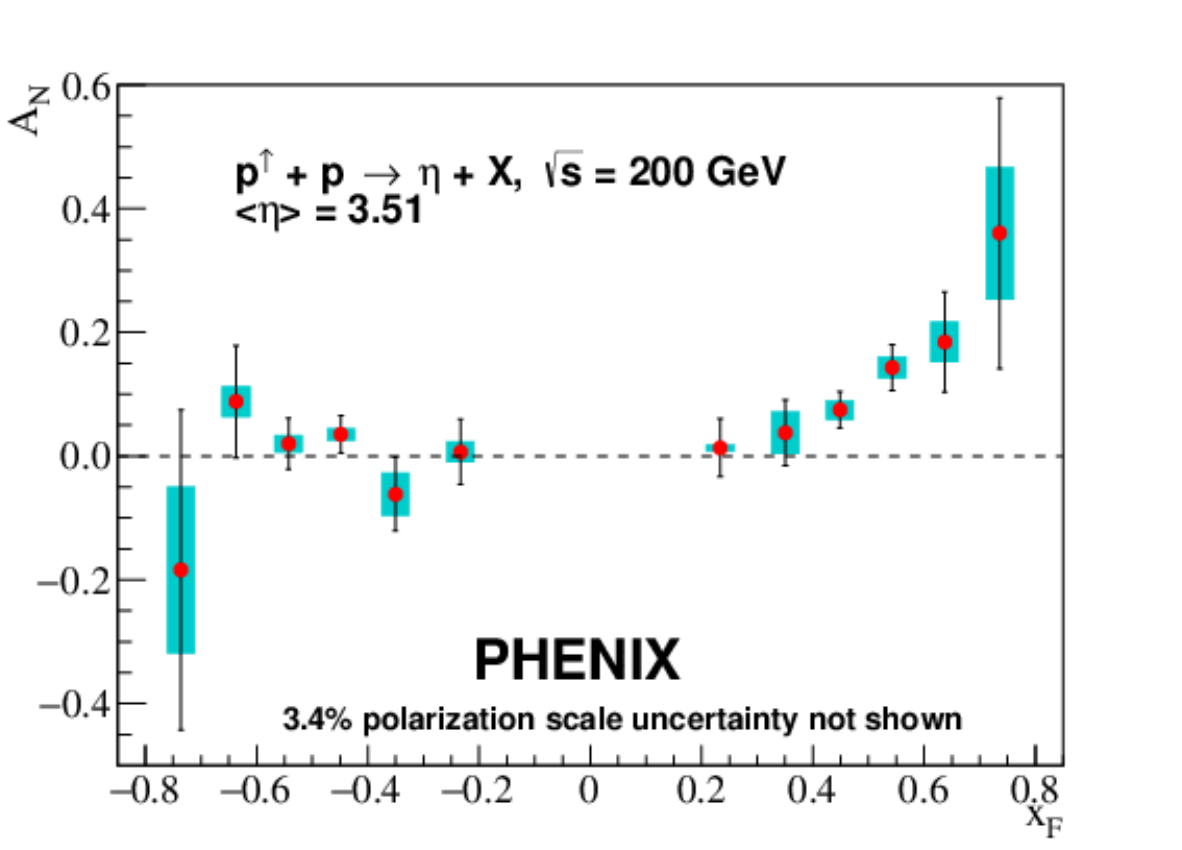}
\caption{\label{fig:ANvsXF} The $x_F$ dependent $A_N$ of forward $\eta$ 
mesons from PHENIX in the 2012 data taking period. The error bars show 
the statistical uncertainty and the error boxes show the uncorrelated 
systematic uncertainties added in quadrature. A 3.4\% global 
polarization scale uncertainty is not shown.}
\end{figure}

The sideband regions are chosen to be sufficiently far from the signal 
region to remove nearly all $\eta \rightarrow \gamma \gamma$ pairs but 
close enough to the peak to sample a similar composition of the 
background under the peak. The low-mass sidebands are defined at 
$0.32<M_{\gamma\gamma}<0.42$~GeV/$c$$^{2}$ and the high-mass sidebands are 
defined at $0.86<M_{\gamma\gamma}<0.96$~GeV/$c$$^{2}$ 
($0.76<M_{\gamma\gamma}<0.86$~GeV/$c$$^{2}$) in the MB (MPC-triggered) data, 
where the MB high sideband is shifted higher to avoid the 
$\omega$(782) region. In both sidebands, $A_N^{\rm bkg}$ is consistent with 
zero within statistical uncertainties for all $x_F$ and $p_T$ bins.

\subsection{\label{sec:systs_AN}Systematic Uncertainties}

Three sources of systematic uncertainty on the TSSAs, all of which are 
uncorrelated across $x_F$ and $p_T$, are added in quadrature for the 
total systematic uncertainty in each $x_F$ or $p_T$ bin. The difference 
in $A_N$ when using the two raw asymmetry extraction methods is taken as 
a systematic and ranges from 0.005--0.060. Another systematic uncertainty 
comes from the difference between background fraction estimation when 
using functional fits and Gaussian-process regression($<$0.01--0.03). Any 
remaining systematic uncertainties are estimated by deviations of the 
asymmetries away from what could be explained by their statistical 
uncertainty. This is accomplished by randomizing the bunch-by-bunch 
polarization directions of the protons and then recalculating $A_N$. 
This bunch shuffling technique removes any knowledge of proton 
polarization and should generate asymmetries consistent with zero. Any 
artificial shift or unaccounted for systematic uncertainty can be seen 
by generating a sample of 10,000 bunch shuffled asymmetries and fitting 
$A_N/\sigma_{A_N}$ to a Gaussian. This Gaussian should follow a standard 
normal distribution, $\mathcal{N}(\mu=0,\sigma=1)$. The resulting 
distributions are found to be consistent with $\mu = 0$ (consistent with 
no artificial asymmetry shift) for all $x_F$ and $p_T$ bins. However, 
all $|x_F| > 0.4$ bins and all $p_T$ bins except for $1.0 < p_T < 1.5$ 
and $2.0 < p_T < 2.5$ GeV/$c$ have distributions with widths exceeding 
unity by more than twice their fit uncertainty. These bins are 
assigned a bunch shuffling systematic uncertainty taken as what is 
necessary to be added in quadrature with the statistical uncertainty to 
account for this larger width. The bunch shuffling uncertainties range 
from 0.01--0.10.

\begin{figure}[hb!]
  \includegraphics[width=1.0\linewidth]{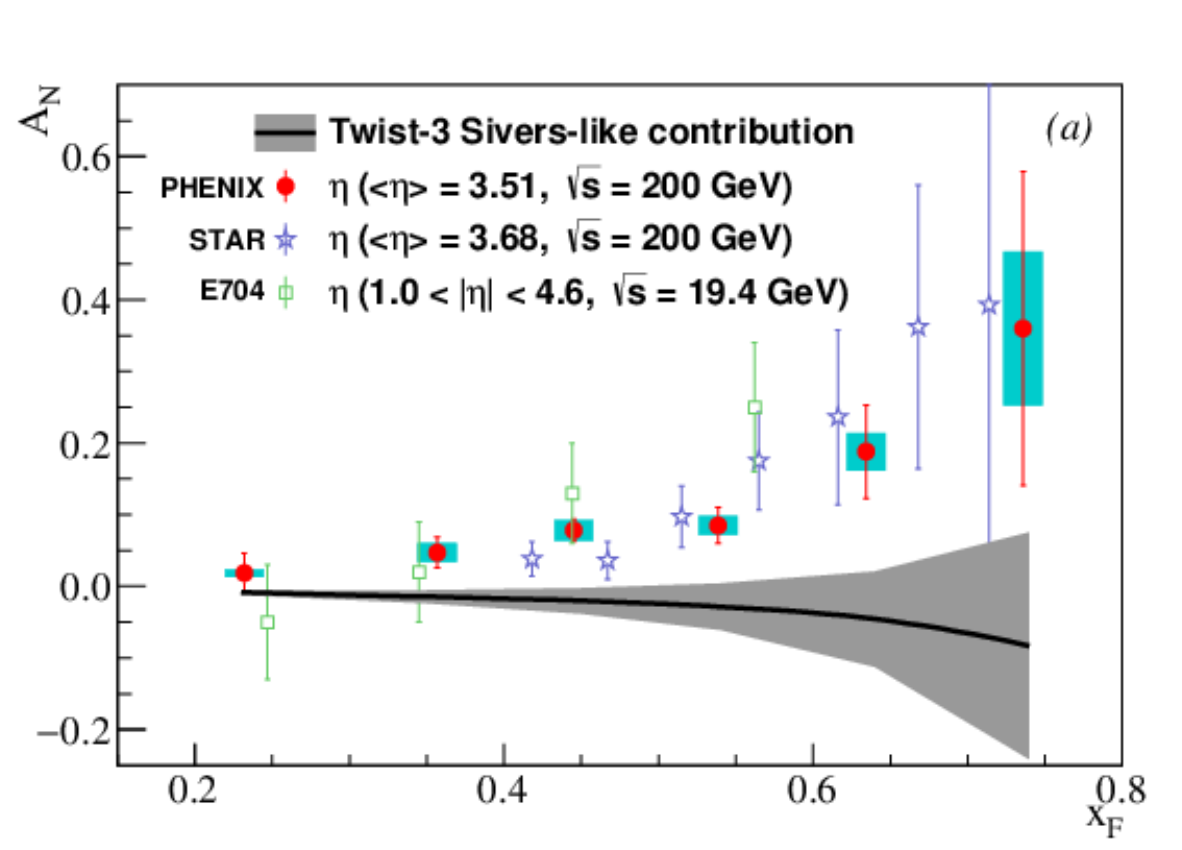}
  \includegraphics[width=1.0\linewidth]{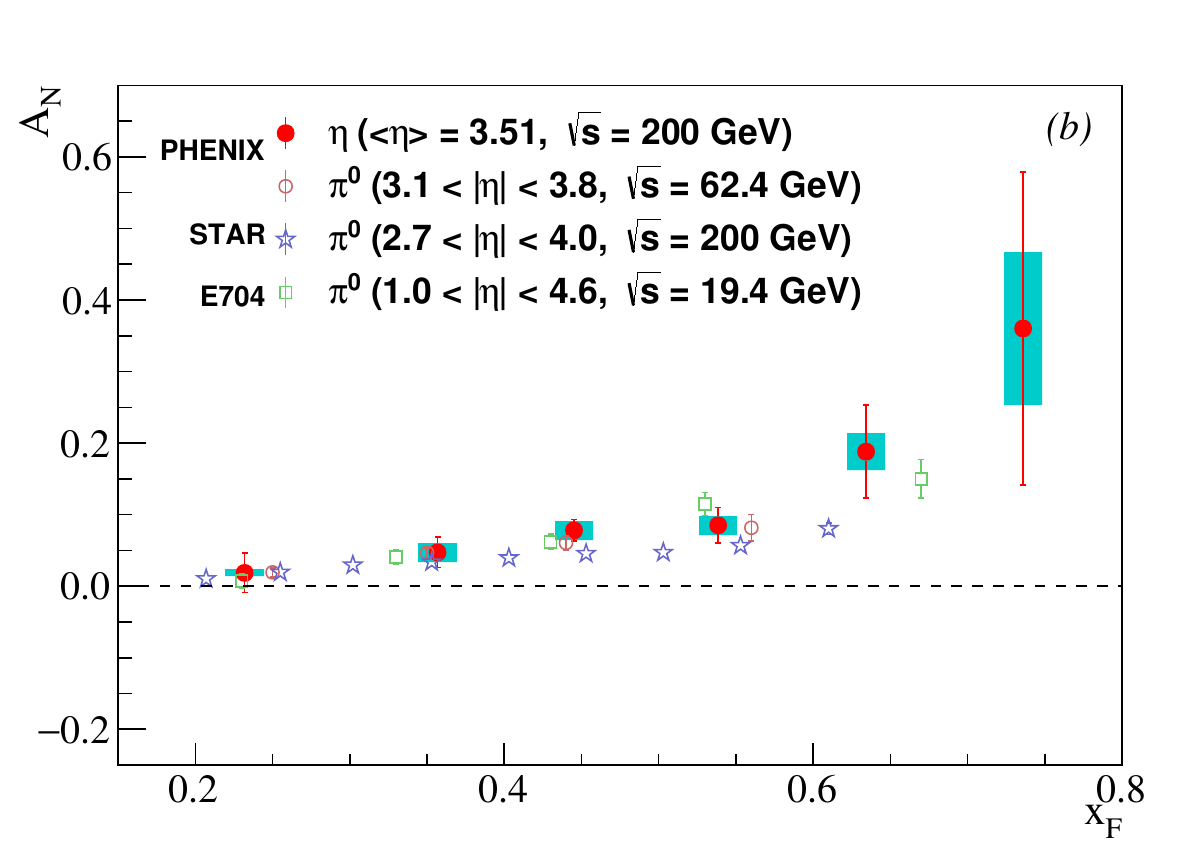}
  \caption{The combined PHENIX measurement of the forward $\eta$ meson 
TSSA from 2008~\protect\cite{PhysRevD.90.072008} and 2012 data. 
Panel (a) shows the asymmetry compared to measurements from 
STAR~\protect\cite{PhysRevD.86.051101} and 
FNAL-E704~\protect\cite{ADAMS19983}, 
and a twist-3 calculation of the initial-state Sivers-like contribution 
to the asymmetry~\protect\cite{Pitonyak_priv,PhysRevD.106.034014}. Panel (b) 
shows the asymmetry compared to previous $\pi^0$ TSSAs from
PHENIX~\protect\cite{PhysRevD.90.012006}, 
STAR~\protect\cite{STAR:2020nnl}, and 
FNAL-E704~\protect\cite{ADAMS1991201}. 
The TSSA at $x_F < 0$ is consistent with zero in both 2008 and 2012 (not 
shown here).}
  \label{fig:ANvsXF_comps}
\end{figure}

\begin{figure}[htb!]
    \includegraphics[width=1.0\linewidth]{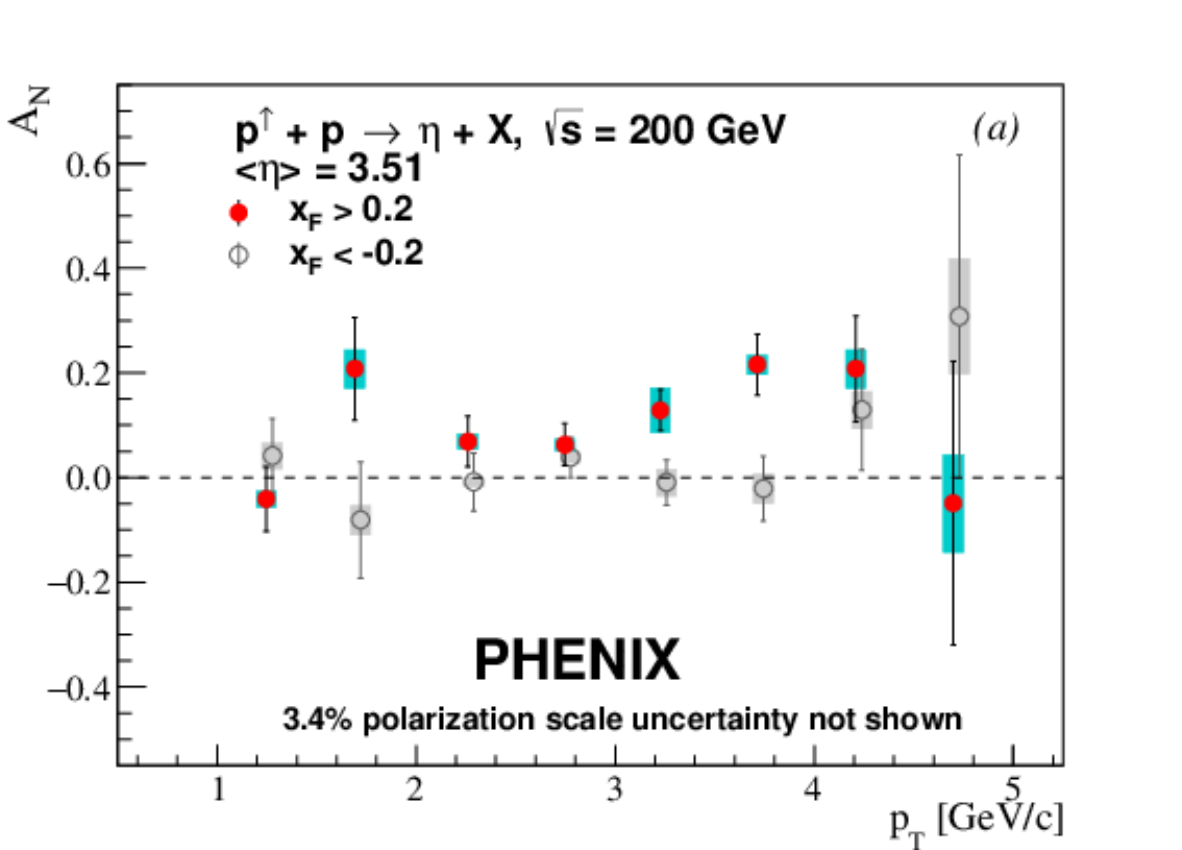}
    \includegraphics[width=1.0\linewidth]{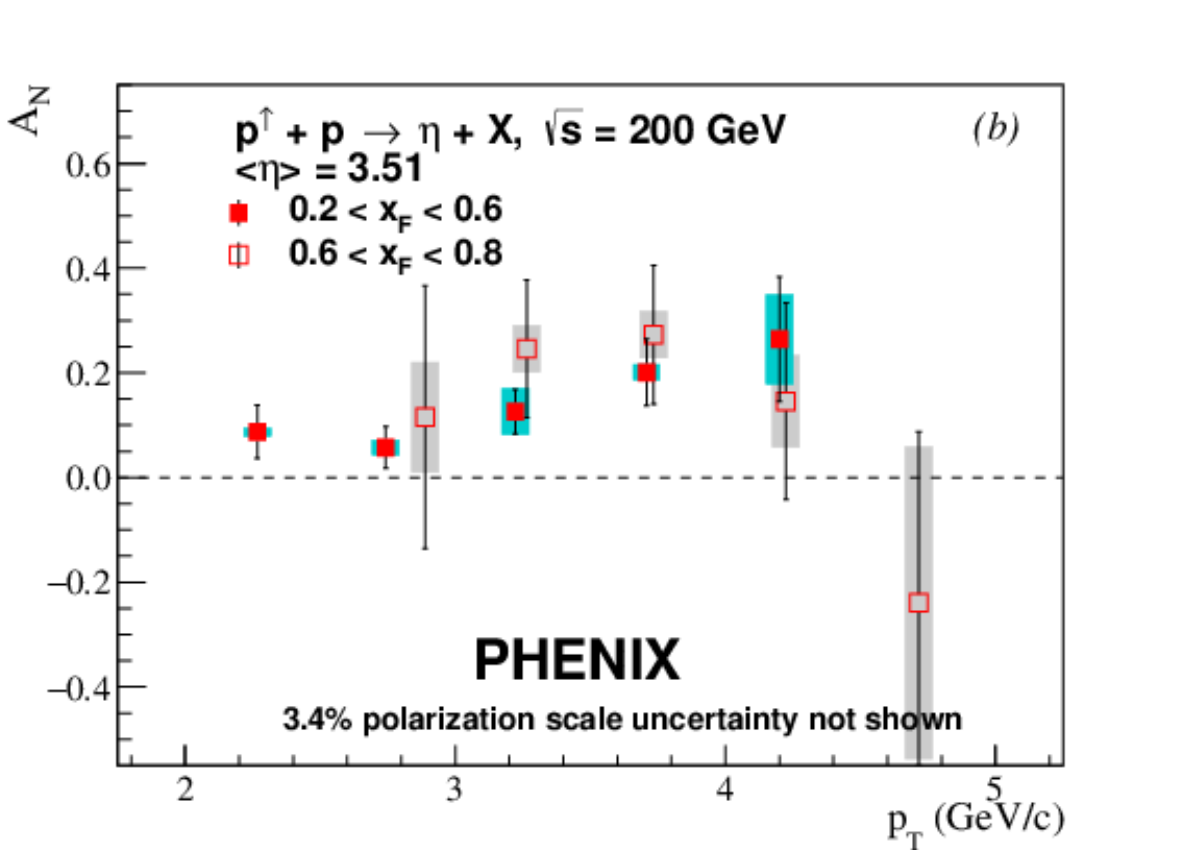}
    \caption{\label{fig:ANvsPT}The $p_T$ dependent $A_N$ of forward 
$\eta$ mesons from PHENIX 2012 data. The error bars show the statistical 
uncertainty and the error boxes show the uncorrelated systematic 
uncertainties added in quadrature. A 3.4\% global polarization scale 
uncertainty is not shown. Panel (a) shows $A_N$ at positive and negative 
$x_F$ while panel (b) shows $A_N$ at positive $x_F$ split into 
intermediate-$x_F$ and high-$x_F$.}
  \label{fig:ptdep2012}
\end{figure}

\section{\label{sec:results}Results and discussion}

Figure~\ref{fig:ANvsXF} shows the $x_F$-dependent $A_N$ of forward 
$\eta$ mesons at $\sqrt{s} = 200$ GeV from the 2012 data taking period. 
The asymmetry at positive $x_F$ increases with $x_F$ up to $\approx$~35\% 
in the highest $x_F$ region above 0.7. The asymmetry at negative $x_F$ 
is consistent with zero within 0.86$\sigma$ of the statistical 
uncertainty.

The combined results from 2008 and 2012 PHENIX data in 
Fig.~\ref{fig:ANvsXF_comps} provide the most precise measurement of the 
forward $\eta$ meson $A_N$ to date. Both the 2012 asymmetry from this 
measurement and combined 2008 + 2012 asymmetry are listed in 
Table~\ref{tab:table1}. Previous $A_N$ measurements from STAR and 
FNAL-E704~\cite{PhysRevD.86.051101,ADAMS19983} are consistent with this 
result in a similar kinematic regime, as seen in 
Fig.~\ref{fig:ANvsXF_comps}(a). A prediction of the twist-3 
initial-state contribution to the asymmetry~\cite{Pitonyak_priv}, 
utilizing the Sivers-like correlator from 
Ref.~\cite{PhysRevD.106.034014} and the newly updated collinear $\eta$ 
meson fragmentation functions~\cite{Aidala:2025kep}, significantly 
underestimates the data. 
Presently, a prediction for the contribution from the 
Collins-like correlator is unavailable for $\eta$ mesons because of the 
lack of SIDIS and $e^{+}e^{-}$ data to constrain the Collins TMD FF.

The $\eta$ meson $A_N$ is also compared to the $\pi^{0} A_N$ from 
PHENIX~\cite{PhysRevD.90.012006}, STAR~\cite{STAR:2020nnl}, and 
FNAL-E704~\cite{ADAMS1991201} in Fig.~\ref{fig:ANvsXF_comps}(b). 
The compatibility of the $\eta$ and $\pi^{0} A_N$ seems to 
indicate that differences in strange-quark content, isospin, or 
mass have a limited impact on the generation of the asymmetry.

The $p_T$-dependent $A_N$ from 2012 data is presented in 
Fig.~\ref{fig:ANvsPT}(a) and Table~\ref{tab:table2}. At positive $x_F$, 
the asymmetry is nonzero and increasing with $p_T$ until $\approx$~4 
GeV/$c$ at which point it levels out. At negative $x_F$, the asymmetry 
is consistent with zero within statistical uncertainty. The high-$p_T$ 
behavior is explored further by splitting the asymmetry into an 
intermediate-$x_F$ ($0.2<x_F<0.6$) and a high-$x_F$ ($0.6<x_F<0.8$) 
region. As seen in Fig.~\ref{fig:ANvsPT}(b), the asymmetry shows the 
flattening or potential suppression at high $p_T$ in the high-$x_F$ 
region that is expected from twist-3 
calculations~\cite{PhysRevD.82.034009,Kanazawa:2011bg,PhysRevD.89.111501} 
and is hinted at in the recent STAR forward $\pi^0$ 
results~\cite{PhysRevD.103.072005}.

\section{\label{sec:summary}Summary}

The transverse single-spin asymmetries of $\eta$ mesons at forward 
rapidity in $\sqrt{s}=200$ GeV $p^{\uparrow}+p$ collisions have been 
measured by the PHENIX collaboration. The TSSAs are measured as a 
function of $\eta$ meson $x_F$ and $p_T$. Previous measurements of large 
asymmetries in the forward region are confirmed with greater precision 
at high $x_F$. Twist-3 predictions of the TSSA using only the 
initial-state Sivers-like correlator do not fully describe these 
asymmetries and comparisons between the $\eta$ and $\pi^0$ TSSAs 
indicate that effects on the asymmetry that could be caused by mass, 
isospin, or strange quarks are relatively minor. At high $x_F$ and 
increasing $p_T$, the $\eta$ meson $A_N$ is consistent with the flat or 
decreasing behavior that has been seen in previous RHIC results and 
twist-3 predictions of light meson $A_N$.

\begingroup \squeezetable
\begin{table*}[htbp!]
\caption{\label{tab:table1}The $A_N$ of forward $\eta$ mesons as a 
function of $x_F$ at $\sqrt{s}=200$ GeV using 2012 PHENIX data, as 
shown in Fig.~\ref{fig:ANvsXF}, and combined 2008+2012 data, as shown 
in Fig.~\ref{fig:ANvsXF_comps}. A global 3.4\% polarization scale 
systematic uncertainty is not included.}
\begin{ruledtabular} \begin{tabular}{ccccccc}
Data Year & $x_F$ bin & $\left<x_F\right>$ & $\left<p_T\right>$ [GeV/$c$] & $A_N [10^{-3}]$ & $\sigma_{stat}[10^{-3}]$ & $\sigma_{syst}[10^{-3}]$\\ \hline
2012 & -0.8 to -0.7 & -0.736 & 4.278 & -184 & 259 & 134 \\ 
     &  -0.7 to -0.6 & -0.637 & 3.717 & 88 & 91 & 24 \\ 
     &  -0.6 to -0.5 & -0.543 & 3.250 & 20 & 41 & 13 \\ 
     &  -0.5 to -0.4 & -0.448 & 2.787 & 35 & 31 & 9 \\ 
     &  -0.4 to -0.3 & -0.353 & 2.161 & -62 & 60 & 34 \\ 
     &  -0.3 to -0.2 & -0.234 & 1.374 & 7 & 53 & 17 \\ 
     &  0.2 to 0.3 & 0.234 & 1.374 & 14 & 47 & 5 \\ 
     &  0.3 to 0.4 & 0.353 & 2.161 & 38 & 53 & 34 \\ 
     &  0.4 to 0.5 & 0.448 & 2.787 & 75 & 30 & 15 \\ 
     &  0.5 to 0.6 & 0.543 & 3.250 & 143 & 37 & 18 \\ 
     &  0.6 to 0.7 & 0.637 & 3.717 & 185 & 81 & 33 \\ 
     &  0.7 to 0.8 & 0.736 & 4.278 & 360 & 219 & 107 \\ \\ 
2008+2012 &  0.2 to 0.3 & 0.232 & 1.358 & 19 & 28 & 5 \\ 
          &  0.3 to 0.4 & 0.357 & 2.311 & 47 & 21 & 13 \\ 
          &  0.4 to 0.5 & 0.445 & 2.708 & 78 & 15 & 14 \\ 
          &  0.5 to 0.6 & 0.539 & 3.134 & 85 & 25 & 13 \\ 
          &  0.6 to 0.7 & 0.634 & 3.608 & 188 & 65 & 26 \\ 
          &  0.7 to 0.8 & 0.736 & 4.278 & 360 & 219 & 107 \\ 
\end{tabular} \end{ruledtabular}
\end{table*}
\begin{table*}[htbp!]
\caption{\label{tab:table2}The $A_N$ of forward $\eta$ mesons from 
PHENIX 2012 data as a function of $p_T$ at $\sqrt{s} = 200$ GeV, as 
shown in Fig.~\protect\ref{fig:ANvsPT}. A global 3.4\% polarization scale 
systematic uncertainty is not included.}
\begin{ruledtabular} \begin{tabular}{ccccccc}
$x_F$ & $p_T$ bin [GeV/$c$] &$\left<p_T\right>$ [GeV/$c$] & $\left<x_F\right>$ & $A_N[10^{-3}]$ & $\sigma_{{\rm stat}}[10^{-3}]$ & $\sigma_{{\rm syst}}[10^{-3}]$\\ \hline
$x_F<-0.2$ 
  &  1.0 to 1.5 & 1.247 & -0.229 & 41 & 71 & 25 \\ 
  &  1.5 to 2.0 & 1.691 & -0.264 & -81 & 111 & 27 \\ 
  &  2.0 to 2.5 & 2.259 & -0.407 & -8 & 55 & 9 \\ 
  &  2.5 to 3.0 & 2.746 & -0.459 & 39 & 38 & 10 \\ 
  &  3.0 to 3.5 & 3.227 & -0.503 & -10 & 44 & 25 \\ 
  &  3.5 to 4.0 & 3.714 & -0.543 & -21 & 62 & 28 \\ 
  &  4.0 to 4.5 & 4.209 & -0.583 & 129 & 116 & 35 \\ 
  &  4.5 to 5.0 & 4.699 & -0.625 & 308 & 309 & 110 \\ \\
$x_F>0.2$ 
  &  1.0 to 1.5 & 1.247 & 0.229 & -41 & 62 & 15 \\ 
  &  1.5 to 2.0 & 1.691 & 0.264 & 208 & 98 & 37 \\ 
  &  2.0 to 2.5 & 2.259 & 0.407 & 69 & 49 & 14 \\ 
  &  2.5 to 3.0 & 2.746 & 0.459 & 63 & 40 & 11 \\ 
  &  3.0 to 3.5 & 3.227 & 0.503 & 129 & 40 & 42 \\ 
  &  3.5 to 4.0 & 3.714 & 0.543 & 216 & 58 & 19 \\ 
  &  4.0 to 4.5 & 4.209 & 0.583 & 208 & 101 & 37 \\ 
  &  4.5 to 5.0 & 4.699 & 0.625 & -50 & 271 & 93 \\ \\
$0.2<x_F<0.6$ 
  &  2.0 to 2.5 & 2.267 & 0.407 & 87 & 51 & 8 \\ 
  &  2.5 to 3.0 & 2.743 & 0.455 & 58 & 40 & 14 \\ 
  &  3.0 to 3.5 & 3.222 & 0.483 & 126 & 42 & 44 \\ 
  &  3.5 to 4.0 & 3.708 & 0.508 & 202 & 64 & 14 \\ 
  &  4.0 to 4.5 & 4.200 & 0.534 & 264 & 119 & 85 \\ \\
$0.6<x_F<0.8$ 
  &  2.5 to 3.0 & 2.888 & 0.619 & 115 & 251 & 104 \\ 
  &  3.0 to 3.5 & 3.264 & 0.640 & 246 & 132 & 44 \\ 
  &  3.5 to 4.0 & 3.734 & 0.656 & 273 & 133 & 44 \\ 
  &  4.0 to 4.5 & 4.224 & 0.666 & 146 & 188 & 87 \\ 
  &  4.5 to 5.0 & 4.716 & 0.671 & -239 & 326 & 298 \\ 
\end{tabular} \end{ruledtabular}
\end{table*}
\endgroup

\section*{ACKNOWLEDGMENTS}

We thank the staff of the Collider-Accelerator and Physics Departments at 
Brookhaven National Laboratory and the staff of the other PHENIX 
participating institutions for their vital contributions. We also thank D. 
Pitonyak for fruitful discussions. 
We acknowledge support from the Office of Nuclear Physics in the
Office of Science of the Department of Energy,
the National Science Foundation,
Abilene Christian University Research Council,
Research Foundation of SUNY, and
Dean of the College of Arts and Sciences, Vanderbilt University
(U.S.A),
Ministry of Education, Culture, Sports, Science, and Technology
and the Japan Society for the Promotion of Science (Japan),
Natural Science Foundation of China (People's Republic of China),
Croatian Science Foundation and
Ministry of Science and Education (Croatia),
Ministry of Education, Youth and Sports (Czech Republic),
Centre National de la Recherche Scientifique, Commissariat
{\`a} l'{\'E}nergie Atomique, and Institut National de Physique
Nucl{\'e}aire et de Physique des Particules (France),
J. Bolyai Research Scholarship, EFOP, HUN-REN ATOMKI, NKFIH,
and OTKA (Hungary),
Department of Atomic Energy and Department of Science and Technology (India),
Israel Science Foundation (Israel),
Basic Science Research and SRC(CENuM) Programs through NRF
funded by the Ministry of Education and the Ministry of
Science and ICT (Korea).
Ministry of Education and Science, Russian Academy of Sciences,
Federal Agency of Atomic Energy (Russia),
VR and Wallenberg Foundation (Sweden),
University of Zambia, the Government of the Republic of Zambia (Zambia),
the U.S. Civilian Research and Development Foundation for the
Independent States of the Former Soviet Union,
the Hungarian American Enterprise Scholarship Fund,
the US-Hungarian Fulbright Foundation,
and the US-Israel Binational Science Foundation.

\section*{DATA AVAILABILITY}

The data that support the findings of this article are not publicly 
available. The numerical values for data shown in 
Figs.~\ref{fig:ANvsXF} and \ref{fig:ANvsXF_comps} are given in Table~I 
and for data shown in Fig.~\ref{fig:ANvsPT} are given in Table~II. All 
values in the plots associated with this article will be stored in 
HEPData~\cite{hepdata} and a link will be provided in an arXiv update.




%
 
\end{document}